\documentstyle[12pt]{article}
\begin{document}
\newcommand{\newc}{\newcommand}
 
\newc{\be}{\begin{equation}}
\newc{\ee}{\end{equation}}
\newc{\ba}{\begin{eqnarray}}
\newc{\ea}{\end{eqnarray}}
\newc{\bea}{\begin{eqnarray}}
\newc{\eea}{\end{eqnarray}}
\newc{\D}{\partial}
\newc{\ie}{{\it i.e.} }
\newc{\eg}{{\it e.g.} }
\newc{\etc}{{\it etc.} }
\newc{\etal}{{\it et al.} } 
\newc{\ra}{\rightarrow}
\newc{\lra}{\leftrightarrow}
\newc{\no}{Nielsen-Olesen }
\newc{\lsim}{\buildrel{<}\over{\sim}}
\newc{\gsim}{\buildrel{>}\over{\sim}}
 
\begin{titlepage}
\begin{center}
January 1997\hfill       
CRETE-97/10 \\
\hfill DEMO-HEP-97/02 \\
             \hfill    
hep-ph/9702221
\vskip 0.5cm
 
{\large \bf
TOPOLOGICAL DEFECTS WITH NON-SYMMETRIC CORE
}

\vskip .1in
{\large Minos Axenides}\footnote{E-mail address:
 axenides@gr3801.nrcps.ariadne-t.gr },\\[.05in]
 
{\em Institute of Nuclear Physics,\\ N.C.R.P.S. Demokritos \\
153 10, Athens, Greece 
}

\vskip .1in
{\large Leandros Perivolaropoulos}\footnote{E-mail address:
leandros@physics.uch.gr},\\[.05in]
 
{\em Department of Physics\\
University of Crete\\
71003 Heraklion, Greece
}
\end{center}
\vskip .1in
\begin{abstract}
\noindent We demonstrate that  field theories involving explicit
breaking of continous symmetries, incorporate two generic 
classes of topological defects each of which is stable for a particular 
range of parameters.
The first class includes defects of the usual type where the 
symmetry gets restored in 
the core and vacuum energy gets trapped there. 
We show however that these defect solutions
become unstable for certain ranges of parameters and decay not to the 
vacuum but 
to another type of stable defect where the symmetry in {\it not} 
restored in the core. 
In the wall case, initially spherical, bubble-like configurations 
are simulated numerically and shown to evolve generically
towards a {\it planar collapse}.
In the string case, the decay of the {\it symmetric core vortex}
resembles the decay of a semilocal string to a 
skyrmion with the important difference that while the skyrmion is unstable 
and decays to the vacuum,  the resulting {\it non-symmetric vortex}
is topologically stable.
\end{abstract} 
\end{titlepage}

\section{Introduction}

Topological defects\cite{vs94,v85} are stable field configurations 
(solitons\cite{r87}) that arise during phase transitions\cite{k72} in
field theories with spontaneously broken discrete or continous symmetries.
Depending on the topology of the vacuum manifold $M$ they are usually 
identified
as domain walls\cite{v85} (kink solutions\cite{r87}) when
$M=Z_2$, as strings\cite{no73} and one-dimensional textures 
(ribbons\cite{bt94,bt95}) when $M=S^1$, as monopoles 
(gauged\cite{t74,p74,dt80}
or global\cite{bv89,p92c}) and two dimensional textures 
($O(3)$ solitons \cite{bp75,r87})when $M=S^2$ and three dimensional textures \cite{t89}
(skyrmions\cite{s61}) when $M=S^3$. Topological defects appearing in GUT phase transitions and quantum field 
fluctuations produced during an inflationary phase
\cite{g81,kt90,lss96} constistute two physically 
distinct mechanisms
for the production of the primordial fluctuations that gave rise to
galaxies and large scale structure in the universe. 
The macrophysical predictions of topological defects (particularly cosmic strings) for structure formation 
have been studied
extensively\cite{vs94,tb86,p94,l93,b96,zlb96,p94a} and compared with the 
corresponding observations with encouraging results. 

Defects corresponding to all of the above vacua are characterized by a
conserved quantity called the 'topological charge'\cite{r87} 
which gurantees 
their stability. It essentially counts the
number of times the field configuration covers the vacuum manifold of the
theory. This is achieved by the variation of the
field at either spatial infinity or over the whole of space. Thus
topological defects could be classified in two broad categories. In the first
category the topological charge becomes non-trivial due to the behavior of the
field configuration at spatial infinity. In this case continuity, forces the
field to remain out of the vacuum manifold at a localized region in space
where the spontaneously broken symmetry gets restored. This region is the 
{\it core} of the defect and is associated with vacuum trapped potential
energy. Because the symmetric phase of the theory is retained in the core of
these defects we will call them hereafter {\it symmetric defects}. Domain
walls, strings and monopoles belong to this class.

In the second category the vacuum manifold gets covered completely as the field
varies over the whole of coordinate space.
The field variable thus remains in the vacuum at all points of coordinate 
space.
Its value at infinity is moreover identified
with a single point of the vacuum manifold. Textures \cite{t89} 
(skyrmions \cite
{s61}), $O(3)$ solitons \cite{bp75} 
(two dimensional textures \cite{t89,p92}) 
and ribbons 
\cite{bt94} (one-dimensional textures \cite{p92}) belong to this class which we
will call for definitenss {\it 'texture-like'} defects. Thus both the behavior of the
field in the  core of the defect (symmetric or non-symmetric phase) as well as its  behavior
at spatial infinity (covering or not the vacuum manifold) suffice to
classify it as belonging to either the {\it symmetric class} or the {\it 
texture class}.

Even though this classification can incorporate most of the well known
defects it is not difficult to think of configurations that belong to neither
class and yet possess quite interesting properties. These are
configurations, appearing mainly in models with explicit symmetry breaking. There 
the field variable covers the whole vacuum manifold at infinity with the core
remaining in the non-symmetric phase. Thus it is possible to have
domain walls, strings and monopoles with cores in the non-symmetric phase of
the theory. For definiteness we will call these {\it 'non-symmetric'} defects.

Examples of nonsymmetric defects have been discussed previously in
the literature. Everett and Vilenkin \cite{ve82},in particular, pointed out 
the existence of domain
walls and strings with non-symmetric cores. In their model however
the walls (strings) were bounded by strings (monopoles) formed at a
phase transition at a higher energy scale. Thus, these defects were unstable to shrinking and
collapse due to their string tension. More recently, Dvali et.al. 
\cite{dtn95} and Bachas et.al.\cite{bt95} considered a model 
which admits non-symmetric walls. They moreover showed the existence of classically stable
bound states of such walls in this model which they  called '$2\pi$ walls'or membranes. 
They speculated that these states 
could be
cosmologically disastrous even if the $Z_2$ symmetry is anomalous. 
These bound state wall configurations were previously
identified with one dimensional textures \cite{p92} or ribbons \cite{bt94} and are
stabilized either by space compatification or by introducing a small
anomalous term that breaks explicitly the $Z_2$ symmetry\cite{ptww91,bt94}.

A particular case of non-symmetric {\it gauge} defect was recently considered by
Benson and Bucher \cite{bb93} (see also Refs. \cite{pr92,h92}) who pointed out that the
decay of an electroweak semilocal string leads to a gauged 'skyrmion' with non-symmetric
core and topological charge at infinity. This skyrmion however, rapidly expands and
decays to the vacuum.

Further back and in the context of $SU(5)$ GUT monopoles \cite{dt80} 
higher $SU(2)$ embeddings into the
fundamental qunituplet space of $SU(5)$ were shown to correspond to 
'higher strength' monopole configurations \cite{sz83}.
These were shown to be unstable and decay into the 'single-strength' 
fundamental monopoles while preserving 
the overall topological number. 
\par
Our goal in this paper is to present more examples of topological defects that
belong to what we defined as the 'non-symmetric' class. We will study in
detail the properties of the simplest such configurations. More specifically  we
will first consider a model with a $U(1)$ symmetry explicitly broken to $Z_2$ which is inturn spontaneously broken.
We will demonstrate that the model admits all three
types of defects (symmetric domain walls, ribbons and non-symmetric walls)
and identify semi-analytically the range of parameters in which each type of 
domain walls is topologically stable.

In an attempt to study the cosmological effects of $2\pi$ walls (ribbons) we
perform numerical simulations of collapsing bubbles of symmetric and
non-symmetric walls. We do not see evidence of the formation of bound states 
($ 2\pi$ walls). Instead, in the non-symmetric parameter range, the bubbles
collapse asymmetrically (pancake collapse) and decay into the vacuum after a
wall collision accelerated by the tension of the bubble. In the symmetric
parameter range (symmetric phase in the wall core) the collapse is spherical
and the bubble decays to the vacuum due to tension as expected. We give
qualitative reasoning for this bubble behavior.

In section 3 we extend our analysis to higher dimensional defects and in
particular to the case of non-symmetric global strings. We identify the
parameter ranges for stability of symmetric and non-symmetric strings in a
theory with a global $SU(2)$ symmetry explicitly broken to $U(1)$.
These results are then extended to the case of the semilocal string
\cite{va91,bv94,p94b,akpv92} where we
identify the parameter range for the symmetric and non-symmetric phase. This
result is in agreement with previous analyses\cite{h92a,akpv92,l92}. We also briefly
discuss the possibility of gauging the configurations considered which could
lead to gauged strings and monopoles with non-symmetric core and thus to
possible generalizations of electroweak strings\cite{n77,v92,jpv92,jpv93}
and monopoles.

Finally, in section 4 we conclude, summarize and discuss the outlook of this
work.

\section{Domain Walls: Symmetric vs NonSymmetric Core}

Consider a model with a $U(1)$ symmetry explicitly broken to a $Z_2$. This
breaking can be realized by the Lagrangian density \cite{vs94,dtn95} 
\begin{equation}
{\cal {L}}={1\over 2}\partial _\mu \Phi ^{*}\partial ^\mu \Phi +
{Ì^2 \over 2}|\Phi |^2 + {m^2 \over 2}Re(\Phi ^2) - {h \over 4}|\Phi |^4
\end{equation}
where $\Phi =\Phi _1+i\Phi _2$ is a complex scalar field. After a rescaling 
\begin{eqnarray}
\Phi &\rightarrow &{m\over \sqrt{h}} \Phi \\
x  & \rightarrow & {1\over m} x \\
M & \rightarrow & \alpha m 
\end{eqnarray}
the potential takes the form
\be
V(\Phi) = -{m^4 \over {2
h}} (\alpha^2 |\Phi|^2 + Re(\Phi^2) - {1\over 2} |\Phi|^4)
\ee
For $\alpha
< 1$ this is a saddle point potential i.e. at $\Phi = 0$ there is a local
minimum in the $\Phi_2$ direction but a local maximum in the $\Phi_1$ (Fig
1). For $\alpha > 1$ the local minimum in the $\Phi_2$ direction becomes a
local maximum but the vacuum manifold remains disconnected, and the $Z_2$
symmetry remains. This type of potential may be called a 'Napoleon hat' 
potential in
analogy to the Mexican hat potential that is obtained in the limit $\alpha
\rightarrow \infty$ and corresponds to the restoration of the $S^1$ vacuum
manifold.

\begin{figure}
\begin{center}
\unitlength1cm
\begin{picture}(6,3)
\put(-3.5,-6.5){\includegraphics{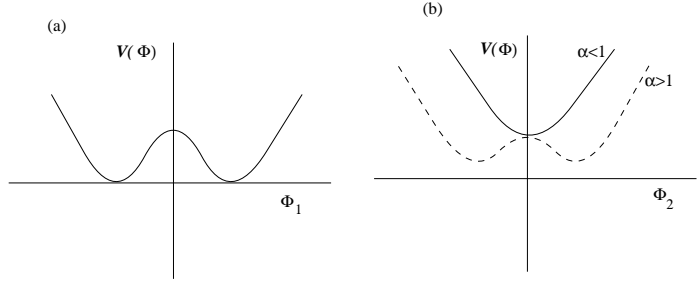}}
\end{picture}
\end{center}
\caption{(a) The domain wall potential has a local maximum
at $\Phi = 0$ in the $\Phi_1$ direction. (b) For $\alpha > 1$ ($\alpha < 1$)
this point
is a local maximum (minimum) in the $\Phi_2$ direction.
}
\end{figure}

\par
The corresponding equation of motion for the field $\Phi$
is
\be
{\ddot \Phi} - {\nabla ^2} \Phi - (\alpha^2 \Phi + \Phi^*) -
|\Phi|^2 \Phi = 0
\ee
which accepts the static kink solution
\bea
\Phi_1
&=& \Phi_R \equiv \pm (\alpha^2 + 1)^{1/2} \tanh (({{\alpha^2 + 1}\over 2})^{1/2}
x) \\
\Phi_2 &=& 0
\eea
This solution corresponds to a {\it symmetric} domain
wall since the core of the soliton is symmetric ($\Phi(0) = 0$) and the
topological charge is resulting from the behavior of the field at infinity 
($Q={1 \over 2}(\Phi(-\infty) - \Phi(+\infty))/ (\alpha^2 + 1)^{1/2} $). The form of the
potential however implies that the symmetric wall solution may not be stable
for $\alpha >1$ since in that case the potential energy favors a solution
with $\Phi_2 \neq 0$. However, the answer is not obvious because for $\alpha
> 1$, $\Phi_2 \neq 0$ would save the wall potential energy but would cost
additional gradient energy as $\Phi_2$ varies from a constant value at $x=0$
to 0 at infinity. A stability analysis is therefore needed and may be
performed as follows.
\par
Consider an ansatz of the form
\be
\Phi =
\Phi_R (x) + \delta \Phi_1 (x) e^{i \omega_1 t} + i \delta \Phi_2 (x)
e^{i\omega_2 t}
\ee
Substituting (9) in (6) we obtain to first order in the
perturbations
\bea
-\delta  \Phi_1^{\prime \prime} & +  & 3 \tanh^2 ({x\over 
\sqrt{2}}) \delta \Phi_1 - \delta \Phi_1 = {\omega_1^2 \over {\alpha^2 +1}}
\delta \Phi_1 \\
-\delta \Phi_2^{\prime \prime} & +  & \tanh^2 ({x\over \sqrt{2}}
) \delta \Phi_2 - {{\alpha^2 - 1} \over {\alpha^2 + 1}} \delta \Phi_2 =
{\omega_2^2 \over {\alpha^2 +1}} \delta \Phi_2 
\eea
For stability we demand
that $\omega_i^2 > 0$ i.e that there are no negative eigenvalues to the
Schroedinger-like equations (10) and (11). The same conditions are obtained by
perturbing the static energy functional
\be
E = {m^4 \over {2 h}}
\int_0^\infty dx ({\Phi^\prime}^2 + {1\over 2} |\Phi|^4 - (\alpha^2 |\Phi|^2
+ Re (\Phi^2))
\ee
to second order around the solution (7), (8) and demanding
that the perturbations contribute no negative part to the energy.
\par
The
potential of the Schroedinger-like equation (11) is positive definite for 
$\alpha < 1$ and it becomes negative and deeper for $\alpha > 1$ as $\alpha$
increases. To find the critical value of $\alpha$ such that for $\alpha >
\alpha_{crit}$ (11) has negative eigenvalues we solve (11) numerically using the
shooting method in the {\it Mathematica} \cite{w91} system. The boundary
conditions at the origin are $\delta \Phi_2 (0) = 1$ and $\delta {
\Phi_2^\prime}(0) = 0$, corresponding to the ground state solution. We solve (11) for
various values of $c\equiv {{\alpha^2 -1} \over {\alpha^2 +1}}$ with $\omega_2
= 0$ looking for $c_{crit}$ such that $\lim_{x\rightarrow \infty} \delta
\Phi_2 (x) = 0$ which corresponds to a ground state with 0 eigenvalue.
Clearly for $c>c_{crit}$ the potential is deeper and there are negative
eigenvalues. Using this method we find $c_{crit} = {1\over 2}$. This solution
corresponds to $\alpha_{crit} = \sqrt{3} \simeq 1.73$. As expected 
$\alpha_{crit} > 1$ due to the effects of the gradient energy discussed
above. Using the same method we also solved equation (10) and found that it
has no negative eigenvalues as expected due to the topological stability of
the component $\Phi_1$.
\par
We have verified this result using two other
methods in order to both test the result and also test the accuracy of the
methods. The first method is based on solving the full non-linear static
field equations obtained from (6) with boundary conditions
\bea
\Phi_1 (0)
&=& 0 \hspace{1cm} \lim_{x\rightarrow \infty} \Phi_1 (x) = (\alpha^2
+1)^{1/2} \\
{\Phi_2^\prime} (0) &=& 0 \hspace{1cm} \lim_{x\rightarrow
\infty} \Phi_2 (x) = 0
\eea

\begin{figure}
\begin{center}
\unitlength1cm
\begin{picture}(6,4)
\put(-3.0,-1.7){\includegraphics{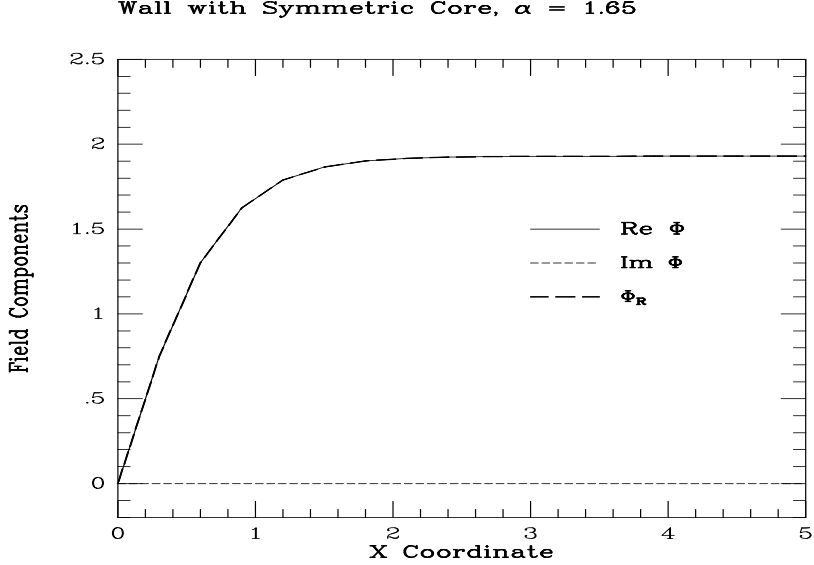}}
\end{picture}
\end{center}
\caption{Field configuration for a symmetric wall with $\alpha = 1.65$.
}
\end{figure}

Using a relaxation method based on collocation
at gaussian points \cite{numrec} to solve the system (6) of second order non-linear
equations we find that for $\alpha < \sqrt{3}$ the solution relaxes to the
expected form of (7) for $\Phi_1$ while $\Phi_2 = 0$ (Fig. 2). For $\alpha > 
\sqrt{3}$ we find $\Phi_1 \neq 0$ and $\Phi_2 \neq 0$ (Fig. 3) obeying the
boundary conditions (13), (14) and giving the explicit solution for the
non-symmetric domain wall. In both cases we also plot the analytic solution
(7) stable only for $\alpha < \sqrt{3}$ for comparison (bold dashed line). As
expected the numerical and analytic solutions are identical for $\alpha < 
\sqrt{3}$ (Fig. 2).

\begin{figure}
\begin{center}
\unitlength1cm
\begin{picture}(6,4)
\put(-3.0,-1.7){\includegraphics{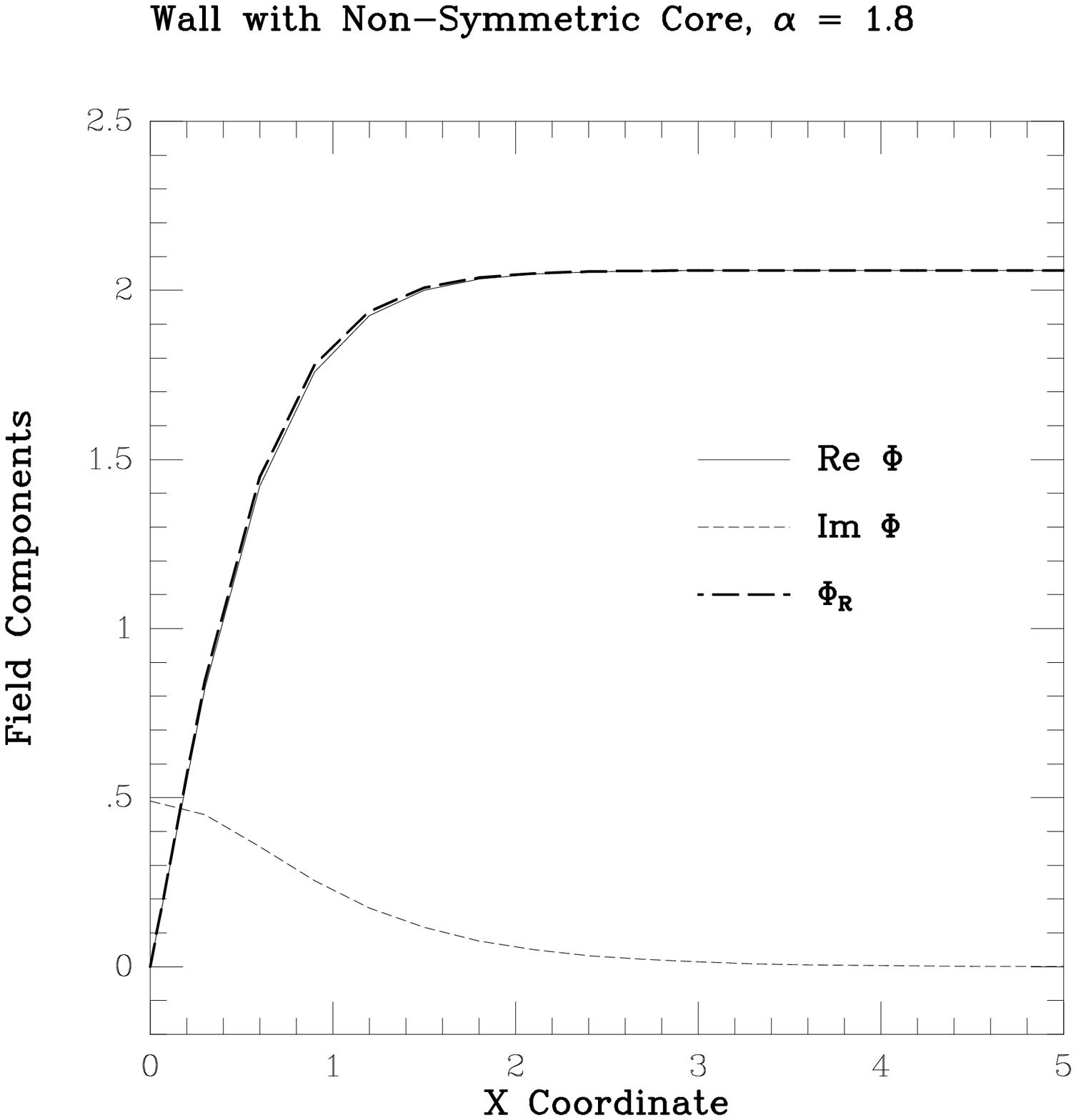}}
\end{picture}
\end{center}
\caption{Field configuration for a non-symmetric wall with $\alpha = 1.8$.
}
\end{figure}

\par
The second method is based on a numerical
minimization of the energy functional (12) using the steepest descent method
of Refs. \cite{bt95,btt96,t96}. We write the energy functional (12) on a one dimensional
lattice of $N$ points and a sum over $2 (N-1) $ variables

\begin{eqnarray*}
E=\sum_{i=1}^{N-1} ({{\Phi_1(i+1)  -  \Phi_1(i)}\over dx})^2  +
({{\Phi_2(i+1)
- \Phi_2(i)}\over dx})^2 + \\ 
{1\over 2} (\Phi_1 (i) ^2 + \Phi_2 (i)^2 )^2 
-\alpha^2 (\Phi_1 (i)^2 + \Phi_2 (i) ^2) - (\Phi_1 (i)^2 - \Phi_2 (i) ^2)
\end{eqnarray*}

We then evaluate numerically the quantities
\be
d\Phi_j (i) \equiv {{dE} \over {d\Phi_j (i)}} 
\hspace{1cm} (j=1,2),\hspace{0.3cm} (i=1,...N)
\ee
and shift $\Phi_j (i)$
to 
\be
\Phi_j (i) = \Phi_j (i) - \epsilon \hspace{0.1cm} d\Phi_j (i)
\ee
where 
\be
\epsilon \simeq 10^{-3} 
\ee
thus finding a new field configuration with lower
energy. We repeat this procedure, keeping the boundary conditions, 
until the energy variation is negligible
implying that we have reached a local minimum in
configuration space. Our initial configuration was the analytic solution (7)
perturbed by a small amount $\Phi_2 = 0.01 \hspace{0.1cm} e^{-x^2}$. 
For $\alpha < \sqrt{3} 
$ the imposed perturbation decreased and the field configuration relaxed to
the analytic solution (7). For $\alpha > \sqrt{3}$ the perturbation increased
towards the solution obtained by the relaxation method (Fig. 3).
The
verification of the result that $\alpha_{cr} = \sqrt{3}$ not only makes it
strongly established but is also a test for the validity of the
methods used.
\par
Bound states of pairs of non-symmetric walls have been
discussed previously in the literature under different names
(one-dimensional textures \cite{p92}, ribbons \cite{bt94}, $2\pi$ walls \cite{dtn95}).
By Derrick's theorem \cite{d64} such configurations are normally unstable
towards expansion induced by the gradient energy term \cite{p92} but can be
stabilized by introducing a small explicit breaking of the $Z_2$ symmetry 
\cite{bt94,dtn95} or by imposing periodic boundary conditions \cite{bt94}.
The
cosmological effects of these states were discussed by Dvali et. al. \cite{dtn95}. 
They pointed out that the stability of a $2\pi$ wall system in the
presence of an anomalous $Z_2$ symmetry can lead to a cosmic overabundance
of walls and therefore restrictions can be imposed on the 
allowed range of parameters. As the cosmic
horizon expands however, the $2\pi$ wall systems will emerge as {\it
bubbles} of non-symmetric walls. It is the evolution and stability of such
bubbles that needs to be studied in order to determine if these objects will
cause a cosmological problem.
\par
In order to address this issue we
constructed a two dimensional simulation of the field evolution of domain
wall bubbles with both symmetric and non-symmetric core. In particular we
solved the non-static field equation (6) using a leapfrog algorithm \cite{numrec}
with reflective boundary conditions. We used an $80 \times 80$ lattice and in all
runs we retained ${{dt} \over {dx}} \simeq {1\over 3}$ thus satisfying the
Cauchy stability criterion for the timestep $dt$ and the lattice spacing $dx$%
. The initial conditions were those corresponding to a spherically symmetric
bubble with initial field ansatz
\be
\Phi (t_i) = (\alpha^2 + 1)^{1/2}
\tanh [({{\alpha^2 +1} \over 2})^{1/2} (\rho - \rho_0)] + 
i \hspace{2mm} 0.1 \hspace{2mm}
e^{- ||x| - \rho_0|} {x \over {|x|}}
\ee
where $\rho = x^2 + y^2$ and $%
\rho_0$ is the initial radius of the bubble. Energy was conserved to within
2\% in all runs. For $\alpha$ in the region of symmetric core stability the
imaginary initial fluctuation of the field $\Phi (t_i)$ decreased and the
bubble collapsed due to tension in a spherically symmetric way as
expected.

\begin{figure}
\begin{center}
\unitlength1cm
\begin{picture}(6,6)
\put(-3.0,-2.0){\includegraphics{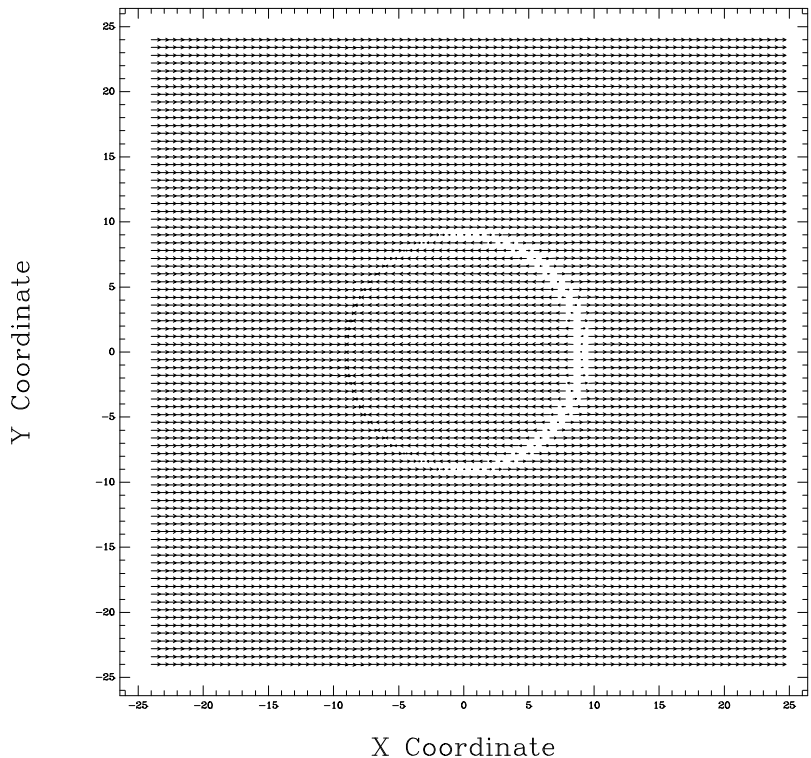}}
\end{picture}
\end{center}
\caption{Initial field configuration for a non-symmetric 
spherical bubble wall with $\alpha = 3.5$.
}
\end{figure}

For $\alpha$ in the region of non-symmetric core stability the
evolution of the bubble was quite different. The initial imaginary
perturbation increased but even though dynamics favored the increase of the
perturbation, topology forced the $Im\Phi (t)$ to stay at zero on two points
along the bubble: the intersections of the bubble wall with the y axis
(Figs. 4, 5). Thus in the region of these points, surface energy (tension)
of the bubble wall remained larger than the energy on other points of the
bubble. The result was a non-spherical collapse with the x-direction of the
bubble collapsing first (Fig. 5). This asymmetric collapse of the bubble
may be understood as follows: The bubble wall in a two dimensional
projection may be approximated by a rectangle with dimensions x-y. Let the x
sides of the rectangle be in the symmetric phase and therefore have energy
per unit length $E_0$. Then the y sides would be in the non-symmetric phase
with energy per unit length $E_1$.

\begin{figure}
\begin{center}
\unitlength1cm
\begin{picture}(6,6)
\put(-3.0,-2.0){\includegraphics{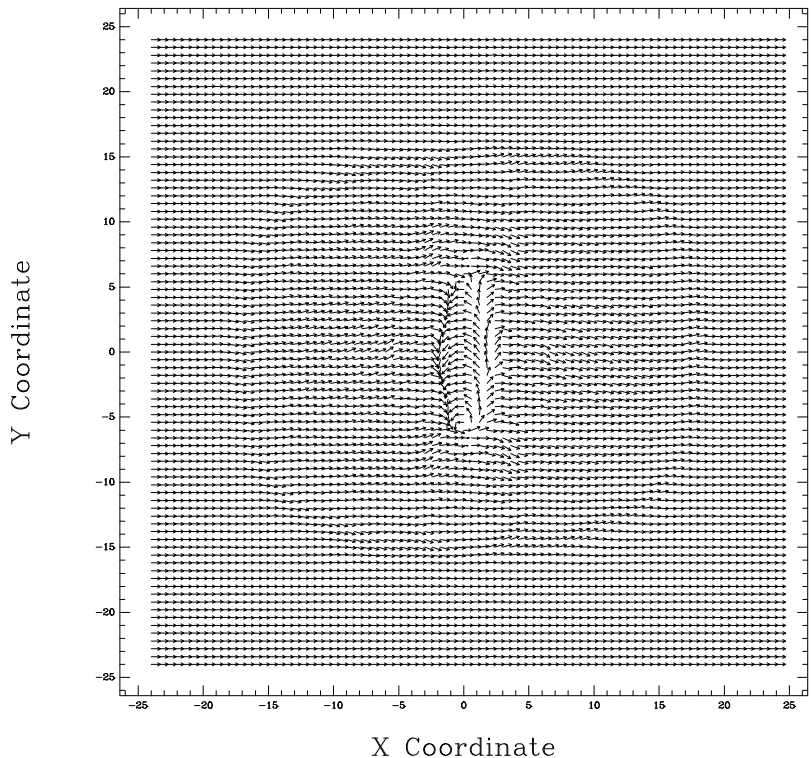}}
\end{picture}
\end{center}
\caption{Evolved field configuration ($t=14.25$, 90 timesteps) for a non-symmetric 
initially spherical bubble wall with $\alpha = 3.5$.
}
\end{figure}

Clearly $E_0 > E_1$ since we are in the
parameter range where the non-symmetric wall is stable. Thus the energy of
the bubble walls is 
\be
E= 2(E_0 x  + E_1 y)
\ee
which implies that it
is energetically favorable for the bubbles to collapse in the x direction
rather than the y direction. This argument clearly does not hold in the
range of symmetric wall stability since in that case the energy would be $E
= 2 E_0 (x + y)$. Also it does not hold for an imaginary perturbation
without the ${x\over{|x|}}$ factor since in that case the energy would be $
E = 2 E_1 (x+y) $. Generically, the two types of perturbations are equally
probable and therefore, in the parameter range of non-symmetric wall
stability we anticipate about half of the wall bubbles to collapse
asymmetrically (pancake collapse) while the rest would collapse in the usual
spherical way. 
\par
In all the cases we simulated we saw no evidence  for
formation of a bound state and in all cases the bubbles collapsed to the
vacuum accelerated by their tension. This decay would be even faster for
an anomalous $Z_2$ symmetry. Thus the cosmological problem discussed in Ref. 
\cite{dtn95} appears not to be realized in the cases we examined.

\section{Generalizations to Other Defects}

It is straightforward to
generalize the analysis of the previous section to higher dimensional
defects like strings or monopoles. Consider for example a model with an $%
SU(2)$ symmetry explicitly broken to $U(1)$. Such a theory is described by
the Lagrangian density:
\be
{\cal{L}} = {1 \over 2} \D_\mu {\Phi^\dagger}
\D^\mu \Phi + {M^2 \over 2} {\Phi^\dagger}\Phi + {m^2 \over 2} {\Phi^\dagger}
\tau_3 \Phi - {h\over 4} ({\Phi^\dagger}\Phi)^2
\ee
where $\Phi =
(\Phi_1, \Phi_2)$ is a complex scalar doublet and $\tau_3$ is the $2 \times 2$
Pauli matrix. After rescaling as in
equations (2)-(4) we obtain the equations of motion for $\Phi_{1,2}$
\begin{figure}
\begin{center}
\unitlength1cm
\begin{picture}(6,4)
\put(-3.0,-1.7){\includegraphics{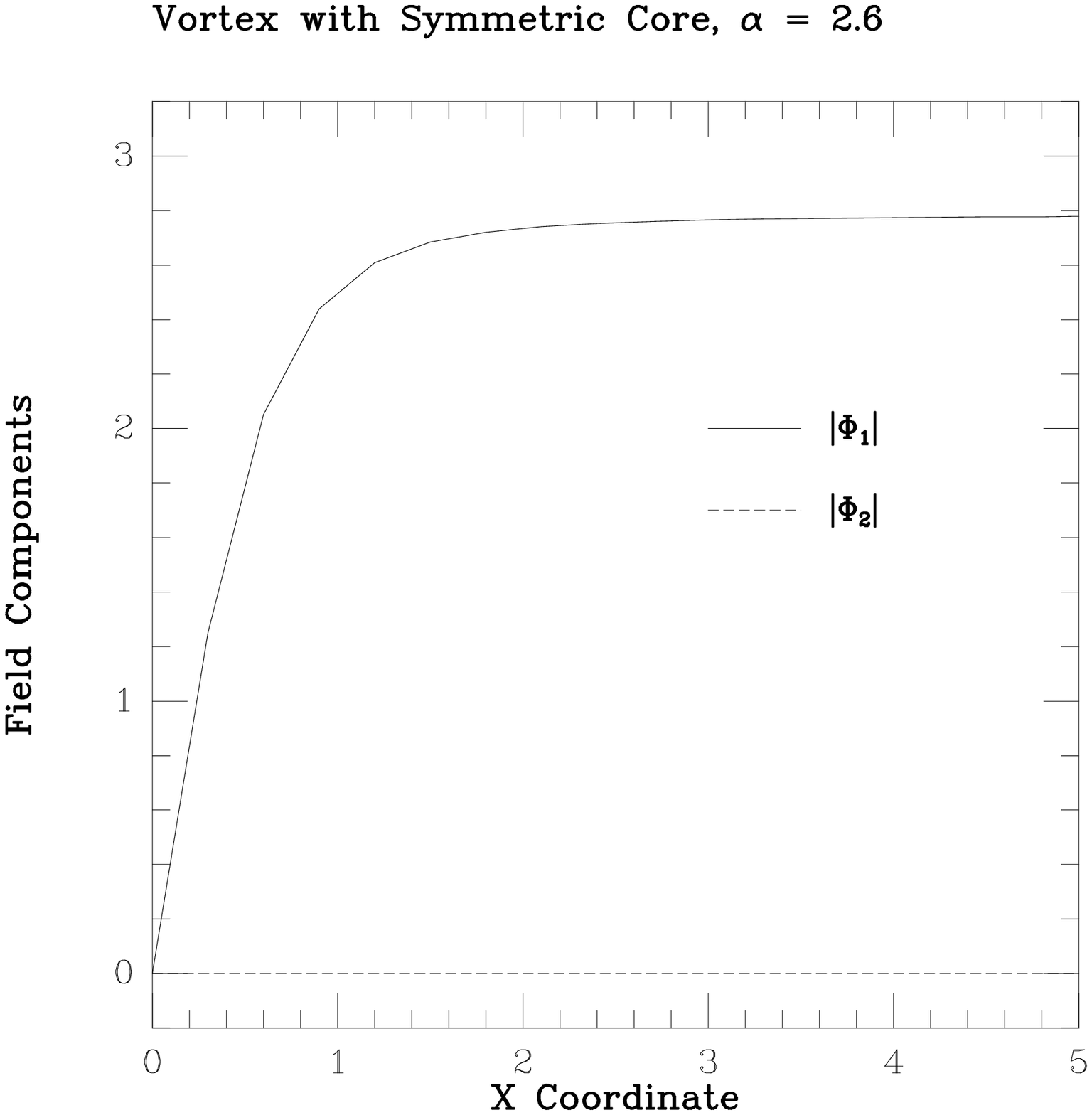}}
\end{picture}
\end{center}
\caption{Field configuration for a {\it symmetric-core}
global string with $\alpha = 2.6$.
}
\end{figure}

\be
\partial_\mu \partial^\mu \Phi_{1,2} - (\alpha^2 \pm 1) \Phi_{1,2}
+({\Phi^\dagger} \Phi) \Phi_{1,2} = 0
\ee
where the + (-) corresponds to
the field $\Phi_1$ ($\Phi_2$).
\par
Consider now the ansatz
\be
\Phi =
\left( \begin{array}{c}
\Phi_1 \\
\Phi_2
\end{array} \right)=
\left( \begin{array}{c}
f(\rho) e^{i\theta} \\
g(\rho)
\end{array}
\right)
\ee
with boundary conditions
\bea
\lim_{\rho \rightarrow 0} f(\rho)
&=& 0, \hspace{3cm} \lim_{\rho \rightarrow 0} { g^\prime} (\rho) = 0
\\
\lim_{\rho \rightarrow \infty} f(\rho) &=& (\alpha^2 +1)^{1/2},
\hspace{1cm} \lim_{\rho \rightarrow \infty} g (\rho) = 0 
\eea

\par

This ansatz
corresponds to a global vortex configuration with a core that can be either
in the symmetric or in the non-symmetric phase of the theory. Whether the
core will be symmetric or non-symmetric is determined by the dynamics of the
field equations. As in the wall case we expect the existence of a critical
value for the parameter $\alpha$ such that for $\alpha < \alpha_{cr}$ the
vortex core is in the symmetric phase ($g(\rho) = 0$) while for $\alpha >
\alpha_{cr}$ the symmetric core configuration is unstable towards decay to a
new vortex configuration with non-symmetric core ($g(0) \neq 0$). To
determine the value of $\alpha_{cr}$  we solved numerically the system (21) of
non-linear complex field equations with the ansatz (22) for various values of the parameter $%
\alpha$. We used the same relaxation technique discussed in the previous
section for the case of walls. For $\alpha < \alpha_{cr} \simeq 2.7$ the
solution relaxed to a lowest energy configuration with $g(\rho) = 0$
everywhere corresponding to a vortex with symmetric core (Fig. 6).

For $
\alpha > \alpha_{cr} \simeq 2.7$ the solution relaxed to a configuration
with $g(0) \neq 0$ indicating a vortex with non-symmetric core (Fig. 7). Both
configurations are dynamically and topologically stable and consist
additional paradigms of the defect classification discussed in the
introduction. 

\begin{figure}
\begin{center}
\unitlength1cm
\begin{picture}(6,4)
\put(-3.0,-1.7){\includegraphics{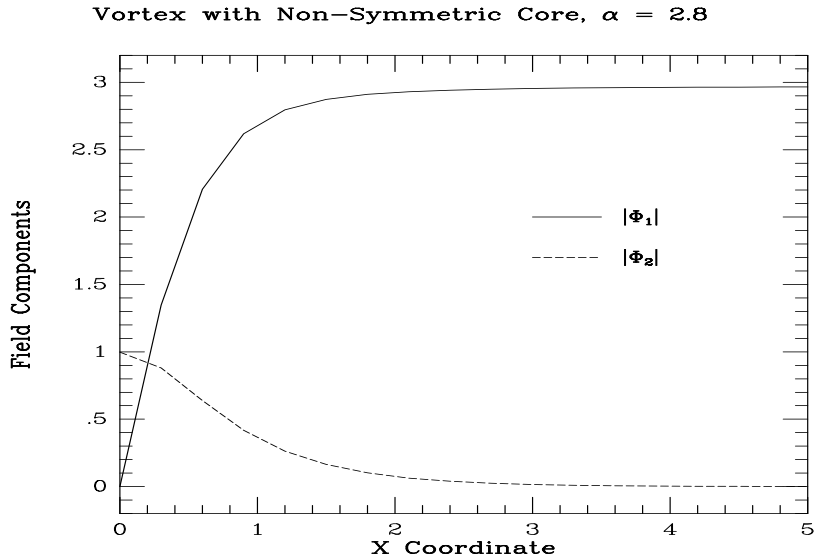}}
\end{picture}
\end{center}
\caption{Field configuration for a {\it non-symmetric-core}
global string with $\alpha = 2.8$.
}
\end{figure}
\par

The existence of a defect with non-symmetric core
corresponding to each 
defect with a symmetric core and winding at infinity
can be established 
for various other cases. For example, our discussion
for global vortices
with non-symmetric core can be trivially generalized to
the case of global
monopoles. Another interesting extension is the case of
non-symmetric 
gauged defects like Nielsen-Olesen vortices or gauge
monopoles. 
Extension of our work in these directions is currently in
progress \cite{approg}.
\par
A common feature of the defects we have discussed
so far is the fact that they
emerge in theories where a larger global symmetry is
partially explicitly broken. This is not always necessary. 
In the case of {\it semilocal strings} a U(1) gauge
symmetry is embedded in a $SU(2)$ global one\cite{va91}.
The Lagrangian density is given by
\be
{\cal{L}}=-{1\over 4} F^{\mu \nu} F_{\mu \nu} + {1\over 2} {
{(D_\mu \Phi )^\dagger}} (D_\mu \Phi) - V(\Phi)
\ee
with $\Phi = (\Phi_1, \Phi_2)$ a complex doublet, 
$D_\mu = \partial_\mu - i e A_\mu$ a $U(1)$ covariant derivative and $V(\Phi) = {\lambda \over 4} 
({\Phi^\dagger} \Phi - \eta^2)^2$ i.e. 
the $SU(2)$ global symmetry is spontaneously broken with the $U(1)$ part of it being gauged.
the $SU(2)$ symmetry is not explicitly 
broken, instead only the $U(1)$ part of it is gauged. This is an alternative
way to single out an $S^1$ subspace of the $S^3$ vacuum manifold.
A similar analysis as the one discussed above (solution of a coupled system 
of non-linear 
ode's by a relaxation method) and an ansatz of the form
(22) with  $A_\mu = {\hat e}_\theta v(\rho)$ leads to a critical value for the
parameter $\beta \equiv {{2 \lambda} \over e^2}$ (the only parameter of the model) such that
for $\beta < \beta_{cr}=1$ the vortex with a symmetric core is stable. 
This result is in agreement with previous studies\cite{h92a,akpv92,l92}.
 For $\beta > \beta_{cr}$ the symmetric core vortex is unstable to a
 configuration with a non-symmetric core called the 'skyrmion' \cite{h92,pr92,bb93}. 
However in this case, since there is no explicit breaking of $SU(2)$,
the 'skyrmion' is not topologically stable (the vacuum is $S^3$ not $S^1$)
and energetics favor expansion and eventual decay to the vacuum.

\section{Conclusion}

We have studied the existence and stability properties of topological
defects with non-symmetric core and non-trivial winding at infinity.
These defects possess cores which are either symmetric or non-symmetric
depending on the range of parameters of the theory. They are 
distinct from previously discussed unstable hybrid\cite{ve82,vs94} defects as they are stable
and are not 
necessarily connected to any other type of defect formed in a 
previous phase transition. 
\par
These results have several interesting implications and potential 
extensions. For example, the observed non-spherical collapse  of 
wall bubbles with non-symmetric core may imply that the domain wall
network simulations need to be re-examined for parameter
ranges where a non-symmetric core in energetically favored. 
\par
Our results are also relevant for baryogenesis mechanisms based on
topological defects\cite{pbdt96,tdb95,bdpt96}. These mechanisms are based on unsuppressed
B+L violating sphaleron\cite{afhkt94} transitions taking place in the 
symmetric core of the defects during scattering processes
\cite{r88,ppdbm91,bdm88,bp88}. 
Thus, the demand for symmetric core in such scenaria
may lead to constraints in the parameters of the corresponding models.
\par
The existense of gauge strings and monopoles with non-symmetric core
is also an interesting extension of our results. Do monopoles exist
with an $SU(3)\times SU(2)\times U(1)$ symmetric core and are they more stable
than the fully symmetric $SU(5)$ one? 
The microphysical and cosmological 
implications of defects with non-symmetric cores is an interesting area
for further investigation.

\section{Acknowledgements}
We are grateful to T. Tomaras and B. Rai for many discussions as well
as for providing us the efficient Fortran routine for energy minimization
of Ref. \cite{bt95}. We are also thankful to C. Bachas for insightful comments
in the initial stages of this work. M.A. acknowledges the hospitality
of the U. of Crete Physics Department.
This work was supported by the EEC grants $CHRX-CT93-0340$ and $CHRX-CT94-0621$ 
as well as by the Greek General Secretariat of Research and Technology grant 
$95E\Delta 1759$.

\end{document}